\documentstyle[11pt]{article}
\begin{document}
\thispagestyle{empty}
\baselineskip 18pt
\rightline{SNUTP-98-113}
\rightline{{\tt hep-th}/9810110}
\rightline{October, 1998} 

\

\def\tr{{\rm tr}\,} \newcommand{\beq}{\begin{equation}}
\newcommand{\eeq}{\end{equation}} \newcommand{\beqn}{\begin{eqnarray}}
\newcommand{\eeqn}{\end{eqnarray}} \newcommand{\bde}{{\bf e}}
\newcommand{\balpha}{{\mbox{\boldmath $\alpha$}}}
\newcommand{\bsalpha}{{\mbox{\boldmath $\scriptstyle\alpha$}}}
\newcommand{\bbeta}{{\mbox{\boldmath $\beta$}}}
\newcommand{\bsbeta}{{\mbox{\boldmath $\scriptstyle\beta$}}}
\newcommand{\blambda}{{\mbox{\boldmath $\lambda$}}}
\newcommand{\bslambda}{{\mbox{\boldmath $\scriptstyle\lambda$}}}
\newcommand{\ggg}{{\boldmath \gamma}} \newcommand{\ddd}{{\boldmath
\delta}} \newcommand{\mmm}{{\boldmath \mu}}
\newcommand{\nnn}{{\boldmath \nu}}

\newcommand{\bra}[1]{\langle {#1}|}
\newcommand{\ket}[1]{|{#1}\rangle}
\newcommand{\sn}{{\rm sn}}
\newcommand{\cn}{{\rm cn}}
\newcommand{\dn}{{\rm dn}}
\newcommand{\diag}{{\rm diag}}

\

\vskip 1cm
\centerline{\Large\bf Sheets of  BPS Monopoles and Instantons}
\vskip 0.2cm
\centerline{\Large\bf with  Arbitrary Simple Gauge Group}
\vskip 1.2cm
\centerline{\large\it
Kimyeong Lee\footnote{Electronic Mail: kimyeong@phya.snu.ac.edu}}
\vskip 3mm 

\centerline{Physics Department and CTP, Seoul National
University, Seoul 151-742, Korea}

\vskip 1.5cm
\centerline{\bf ABSTRACT}
\vskip 2mm
\begin{quote}
{\baselineskip 16pt We show that the BPS configurations of uniform
field strength can be interpreted as those for sheets of infinite
number of BPS magnetic monopoles, and found that the number of
normalizable zero modes per each magnetic monopole charge is four. We
identify monopole sheets as the intersecting planes of D3
branes. Similar analysis on self-dual instanton configurations is
worked out and the number of zero modes per each instanton number is
found to match that of  single isolated instanton. }
\end{quote}


\newpage

There has been some interest in BPS field configuration with uniform
magnetic field~\cite{bak}. This background field exerts no force on
isolated BPS monopoles as the repulsive magnetic force cancels the
attractive Higgs force.  Thus there is a possibility of BPS
configuration of an isolated monopole in such background, which has
been explored recently~\cite{park}. Besides this, there has been a
long standing interest in similar uniform configuration of four
dimensional self-dual equation of Yang-Mills
theories~\cite{leutwyler}. This configuration is known to be stable
and the zero modes of this configuration has been explored to
understand the quantum correction to the effective action.

In this paper we show that the BPS configuration with uniform magnetic
field can be interpreted as a two-dimensional sheet made by infinite
number of BPS magnetic monopoles lying on a plane. By studying zero
modes, we show that the number of normalizable zero modes per unit
magnetic flux is four, identical to that for an isolated
monopole. Along the planar direction of the monopole sheet, the zero
modes are given by the wave function of the lowest Landau level, so
that the monopole sheet appears a quantum phase space or
noncommutative plane.  By considering the case of the general gauge
group which is maximally broken, we show that magnetic monopole sheets
exist for each pair of positive and negative root vectors. Again for
unit magnetic monopole flux, there are four normalizable zero
modes. Especially for SU(N) gauge group, the magnetic monopole sheets
can be identified as intersecting planes of D3 branes in type IIB
string theory. For the instanton case, the zero modes are
well-known. What is new here is that the center position of
normalizable zero modes is pointed out to live on four dimensional
quantum phase space. In theories with general gauge group, we also
show that the number of normalizable zero modes per unit Pontryagin
index is identical to four times the dual Coxeter number, $4c_2(G)$,
which has been obtained from the index theorem for isolated
instantons~\cite{bernard} and from the counting the number of
constituent monopoles of instantons on $R^3\times S^1$~\cite{klee}.

In a Yang-Mills Higgs theory where $SU(2)\rightarrow U(1)$, the field
configuration for  single monopole is spherically symmetric and has a
core region of size $1/m_W=1/v$ with coupling constant $e=1$. Inside
the core the gauge field for the massive $W$ bosons is nonzero, but
vanishes exponentially outside the core.  When we put $N>0$ BPS
magnetic monopoles as close as possible, the core configuration
becomes  complicated. For two monopoles, the core is a
torus~\cite{torus}, for three monopoles, the core is
tetrahedral~\cite{tetra}, and so on.

For large $N$, the shape of the core shape becomes presumably more
involved. Whatever shape it takes, one may ask what is the rough size
of the core region. The answer can be found readily by considering the
asymptotic behavior of the Higgs field.  Since the Higgs field
approaches $(v-N/r)$ asymptotically, the natural scale of the core
size is
\beq
r_{core}\sim N/v=N/m_W.
\label{core}
\eeq

Now imagine configurations with an increasing number of magnetic
monopoles.  A simple way is to put magnetic monopoles along a line
with an equal distance $d$, which, say, is much greater than $1/m_W$.
We know this configuration cannot be axially symmetric as there exits
a relative orientation between any two monopoles. Since the minimum
core size (\ref{core}) is less than the size of the configuration,
monopole cores seem to be separated. However there exits an additional
logarithmic divergence in the monopole core size. To see this,
consider the value of the Higgs field for a fixed point in large $N$
limit. It would be roughly of order
\beq
|\phi| -v \sim \sum_{n=1}^N - \frac{1}{nd} \sim - \frac{1}{d} \ln N.
\eeq
Thus, the core of magnetic monopoles will eventually merge together
when $N\sim e^{dv}$. The large $N$ limit can be taken only if we
increase $v$ simultaneously.  Otherwise, we will end up with just
symmetric phase. This picture is consistent with another view of lined
magnetic monopoles. Far away from the line, the $U(1)$ magnetic field
will fall off like $1/r$ with $r$ being the distance from the
line. This means the Higgs field will increase like $\log r$.

Now imagin putting $N$ monopoles on a two dimensional square lattice
of lattice size $d$. With one monopole on each lattice site, they
occupy a square of size $\sqrt{N}d$. Since the core size (\ref{core})
grows linearly with $N$, the cores of these monopoles will start to
overlap for $N\sim (vd)^2$ and individual monopoles becomes
indistinguishable. As in the linear configuration, we can take the
infinite $N$ limit of the planar configuration only if we take
$v=\infty$ simultaneously. The magnetic field far from the plane is
approximately uniform, which in turn implies that the Higgs field
grows linearly.  There may be several BPS field configurations which
can be interpreted as those corresponding the sheet of BPS magnetic
monopoles.  It appears hard to find any of such BPS configurations.
In this paper, we argue that the well-known homogeneous BPS
configuration
\beq
B^a_i = D_i\phi^a = b \delta_{a3}\delta_{i3}
\label{homo}
\eeq
is one of such configurations.  It is obvious in retrospective that it
should be so. The corresponding Higgs and gauge fields are
\beqn
&& \bar{\phi}^a = b\delta^{a3}(z-z_0), \\
&& \bar{A}^a_i = \frac{b}{2}\delta^{a3} ( -y,x,0)
\eeqn
in the symmetric gauge. Note that the origin of the $x,y$ plane is not
special as there is a translation symmetry on that plane.  However,
the plane $z=z_0$ is very special as the gauge symmetry is restored
only on that plane. The naive $W$ boson mass increases as one moves
away from the plane. Also the gauge invariant $U(1)$ magnetic field
$B_i^a \hat{\phi}^a=b\,{\rm sign}(z-z_0)$ changes the sign as one
crosses the plane.  Both of them implies that the plane could be
interpreted as the sheet of BPS magnetic monopoles.  The individual
characteristics of magnetic monopoles disappear, and so do the
nonabelian characteristics of the core region. Thus there is no core
region at all. Still, there emerges some individuality of monopole as
we will see.  As  single magnetic monopole carries the magnetic flux
$4\pi$ and the magnetic fields of opposite sign come out from the both
sides of the sheet with strength $b$, we can assign the magnetic
monopole number density per unit area to be
\beq
n_{\rm monopole} = \frac{b}{2\pi}.
\label{monod}
\eeq
We note that in the SU(2) case only one  sheet is allowed.

The interesting question is then whether we can take out  single BPS
magnetic monopole away from the sheet. While we do not know the answer
for this question, we can begin by exploring what are the normalizable
zero modes of the magnetic monopole sheet.

The linear fluctuations $\delta A^a_i, \delta \phi^a $ satisfy the
linearized BPS equations 
\beq
 \epsilon_{ijk}(\partial_j \delta A_k^a + \epsilon^{abc}
\bar{A}^b_j\delta A^c_k ) = (\partial_i\delta \phi^a +
\epsilon^{abc}\bar{A}_i\delta \phi^c) - \epsilon^{abc}\bar{\phi}^b
\delta A_i^c,
\label{linear1} 
\eeq
and the background gauge condition
\beq
 (\partial_i \delta A^a_i + \epsilon^{abc} A^b_i \delta A^c_i) +
\epsilon^{abc} \bar{\phi}^b \delta \phi^c = 0.
\label{back1}
\eeq
The background gauge is implied by the Gauss law satisfied by the
slowly moving monopole in the $A_0$ gauge.  The fluctuations $\delta
A^3_i, \delta \phi^3$ along unbroken abelian direction are independent
of the background and non-normalizable. They correspond to the change
in the position and orientation of the magnetic monopole sheet.

For the study of fluctuations along broken symmetry direction, we introduce
\beqn
&& W_i  = \frac{1}{\sqrt{2}} (\delta A^1_i +i \delta A^2_i), \\
&& W_4 = \frac{1}{\sqrt{2}}(\delta \phi^1+i\delta \phi^2),
\eeqn
which describe to massive $W$ bosons in the ordinary case.  The
linearized BPS equations and the background gauge condition become
\beqn
&& \epsilon_{ijk} D_j W_k = D_i W_4- D_4 W_i, \nonumber\\
&& D_i W_i +D_4 W_4=0 , \label{weq}
\eeqn
where
\beqn
D_i = \partial_i -iV_i ,\;\;\;\; V_i = \frac{b}{2} (-y,x,0), \nonumber\\
D_4= \partial_4 - iV_4,\;\;\;\; V_4= b(z-z_0).
\eeqn
Of course the fluctuations considered here are independent of $x^4$ and so
$\partial_4=0$.

These equations can be easily solved. There are two independent families 
of solutions:
\beqn
&& W_1=-i W_2 = c_1 e^{\frac{ib}{2}(x_0y-y_0x) - 
\frac{b}{4}((x-x_0)^2+(y-y_0)^2
-\frac{b}{2}(z-z_0)^2}, \nonumber\\
&& W_3=-i W_4 = c_2 e^{\frac{ib}{2}(x_0y-y_0x) -
 \frac{b}{4}((x-x_0)^2+(y-y_0)^2
-\frac{b}{2}(z-z_0)^2} ,\label{zerom}
\eeqn
where $c_1$ and $c_2$ are two independent constants. Then these two
families and their complex conjugate form four independent families of
zero modes. Along the $z$ direction, the zero modes are given by the
ground state wave function of harmonic oscillator, which is
concentrated on the monopole sheet $z=z_0$. This is due to the fact
that the mass of $W$ boson is zero on that sheet and increases
linearly away from that. Along the $x-y$ direction, the zero modes are
given by the wave function of the lowest Landau level. Since we have
chosen the symmetric gauge for the uniform magnetic field, the wave
function can be chosen to be concentrated on the position $(x_0,y_0)$.

Now let us recall the well known physics of the Landau levels.  The
conserved $x$ and $y$ translations of the wave function do not commute
each other, making the $x-y$ plane a quantized phase space, or
noncommutative plane.  The conserved generators of translations are
$\pi_x=-i\partial_x -eBy/2$ and $\pi_y=-i\partial_y +eBx/2$, satisfying
the commutation relation,
\beq
[\pi_x,\pi_y] = ieB.
\eeq
Thus, the parameters $x_0,y_0$ appearing in the wave function
overcount the number of independent ground wave function. This can be
seen easily by choosing the Landau gauge and compactifying the $x-y$
plane.  The degeneracy is not infinite in spite that $x_0,y_0$ are
continuous. Rather, there is a minimum area $2\pi/eB$ per one state.
Thus, in our notation $e=1,B=b$, the number density of independent
ground states per unit area is given by
\beq
 n_{\rm landau} =\frac{b}{2\pi}.
\eeq
In terms of the magnetic length $l_b=1/\sqrt{b}$, one can understand
the quantum cell $2\pi l_b^2$ as the size of the minimum quantum
volume of the phase plane. Also the total magnetic flux in this
quantum cell is the Dirac quantum flux $2\pi$, which is the minimum
magnetic flux one can put consistently in the compactified $x-y$
space.

Coming back to our magnetic monopole problem, the number density of
the zero modes is four times that of the ground state Landau level,
$4n_{\rm landau} $.  Since the number density of monopole in
Eq.~(\ref{monod}) is identical to the $n_{\rm landau}$, the number of
zero modes per unit magnetic monopole becomes four.
This is exactly identical to the number of zero modes for  single BPS
monopole in SU(2) gauge group. For  one  monopole, three of them
account  its position and one does for its internal phase angle. In our
case monopoles are completely dislocalized on the sheet, but we can
imagine figuratively  single quantum cell as one  monopole.  As
our analysis is at linear level, we do not know exactly how these zero
modes will develop nonlinearly. It would be interesting to find the
full nonlinear distortion of the monopole sheet and the geometry of
the infinite dimensional moduli space of zero modes. Due to the fact
that the $x_0,y_0$ space form noncommutative plane, the moduli space itself
may be noncommutative.

Let us now generalize the above consideration to Yang-Mills Higgs
theory of arbitrary simple gauge group $G$ of rank $r(G)$ and
dimension $d(G)$. We split the generators of the Lie algebra to the
Cartan subalgebra and the raising and lowering operators.  We
normalize the generators so that in the adjoint representation
\beqn 
\tr (H_a H_b)=c_2(G) \delta_{ab} , \label{eigen} \\
\tr(E_\alpha^\dagger E_\beta)=c_2(G)\delta_{\alpha\beta},
\eeqn
where $a,b=1,...,r(G)$.  The normalization factor $c_2(G)$ is the
quadratic Casimir for the adjoint representation and becomes {\it the
dual Coxeter number} when the longest root vectors is normalized to
have length one. (See for example Ref.~\cite{bernard}) Among the
commutation relations, ones we need are
\beq
[{\bf H}, E_\alpha] =\balpha E_\alpha,
\eeq
where ${\bf H}$ and $\balpha$ are $r(G)$ dimensional vectors.  The
eigenvalues of the matrix $H_a H_b$ consist of $\balpha_a\balpha_b$
for all roots and so Eq.~(\ref{eigen}) implies an identity
\beq
\sum_{\balpha}  \balpha_a \balpha_b = c_2(G) \delta_{ab}.
\label{identity}
\eeq

The BPS  equation in this general group becomes
\beq
B_i=D_i \phi,
\eeq
where 
\beqn 
B_i = \frac{1}{2} \epsilon_{ijk}(\partial_j A_k -\partial_k A_j-
i[A_j, A_k]) ,
\\ D_i\phi = \partial_i\phi -i[A_i,\phi]. 
\eeqn 
The uniform solution of the BPS equation can be chosen to lie in the
Cartan subalgebra,
\beq
B_i = D_i\phi={\bf b}_i\cdot {\bf H},
\eeq
where ${\bf b}_i$ are $r$-dimensional vectors for each $i=1,2,3$.  The
corresponding field configuration in the symmetric gauge is 
\beqn
\bar{A_i}= -\frac{1}{2} \epsilon_{ijk}x^j {\bf b}_k\cdot{\bf H},
\nonumber \\
\bar{\phi}= x^i {\bf b}_i \cdot {\bf H} +{\bf h}\cdot {\bf H} .
\label{back3}
\eeqn
There are still remaining gauge transformations: the unbroken
$U(1)^{r(G)}$ transformations and the Weyl reflections which shuffle
the generators in the Cartan subalgebra.	

The linearized BPS equations for $\delta A_i, \delta \phi$ in this
background is
\beq
\epsilon_{ijk}(\partial_j \delta A_k-i [\bar{A}_j,\delta A_k]) = 
\partial_i \delta \phi - i[\bar{A}_i,\delta \phi] -\partial_4 \delta A_i +
 i[\bar{\phi},\delta A_i].
\label{linear3}
\eeq
The background gauge condition is 
\beq
\partial_i\delta A_i -i[\bar{A}_i,\delta A_i] +\partial_4 \delta \phi
 -i[\bar{\phi},\delta \phi]=0.
\label{backg3}
\eeq
Again,  as there is no $x^4$ dependence, $\partial_4=0$.
As in the $SU(2)$ case, the background field $\bar{A}_i, \bar{\phi}$
lies on the Cartan subalgebra and the equations are linear, and so the
linear fluctuations along the generators do not mix.  The fluctuations
$\delta A_i, \delta \phi$ along unbroken gauge groups is independent
of the background fields and unnormalizable. They are associated with
$\delta {\bf b}_i$ and $\delta {\bf h}$.

For the study of fluctuations lying along a raising operator
$E_\balpha$, we  introduce $W_\mu$ such that 
\beq
\delta A_i = W_i E_\alpha, \,\, \delta \phi = W_4 E_\alpha.
\label{wand}
\eeq
The $W_\mu$ equations become identical to those in Eq.~(\ref{weq}) if
we rotate the coordinate so that ${\bf b}_i\cdot \balpha$ points to
$\hat{z}$ direction and put
\beq
b= |{\bf b}_i \cdot \balpha |,\;\;\; z_0 = -{\bf h}\cdot \balpha .
\label{bandz}
\eeq
Then the solution will be identical as those in
Eq.~(\ref{zerom}). Since zero modes  are confined on the plane
\beq
x^i {\bf b}_i \cdot \balpha +{\bf h}\cdot \balpha =0, 
\label{plane}
\eeq
we interpret this plane as the $\balpha$ monopole sheet. Note that
this plane is invariant under $\balpha \rightarrow
-\balpha$. Similarly to the $SU(2)$ case, the number density of
$\balpha$ monopole is
\beq
n_{\balpha} = \frac{|{\bf b}_i\cdot\balpha|}{2\pi},
\eeq
and the number of zero modes per unit quantum cell or BPS monopole is
again four.

When there is unbroken nonabelian subgroup so that the planes
(\ref{plane}) do not overlap, the number of monopole sheet is that of
positive roots $(d(G)-r(G))/2$.  Contrast to the case of finite number
of magnetic monopoles where there is fundamental magnetic monopoles
corresponding to simple roots~\cite{erick}, all monopole sheets in our
case are in equal footing.

The D-brane picture of $N=4$ dimensional Yang-Mills theory in four
dimension helps to understand our result~\cite{polchinski}.  Monopoles
appear as D-strings connecting parallel D3 branes in type 2B string
theory. Following the work by Callan and Maldacena~\cite{callan} these
D-strings can be regarded as continuous deformation of D3 branes. Our
work suggests that as the number of magnetic monopoles increases to
infinite, something drastic happens.  The D3 branes get tilted and
appear to intersecting each other. Since D3 branes are self-dual,
there is no distinction between intersecting and contacting of two D3
branes.  Since the field configuration is self-dual, this
configuration of interacting D3 branes should be supersymmetric. In
the SU(2) case there is an obvious duality between the $z$ direction
and $\phi$ direction in the D-brane picture, but it is not obvious in
the field theory picture. In SU(N) case, we can rewrite the Higgs
field in Eq.~(\ref{back3}) in the $N$ dimensional Hermitian matrix
form. Its $i$-th diagonal component then indicates the position of the
$i$-th D3 brane. Thus, one can identify each monopole sheet associated
with a given root $\balpha={\bf e}_i-{\bf e}_j$ with the plane of
intersection between the $i$-th and $j$-th D3 branes.

Let us now change our attention to the  instanton case. Instantons are
the solutions of self-dual equations,
\beq B_i = F_{i4} \eeq
in Euclidean four dimensional Yang-Mills theory.  The self-dual
configuration with homogeneous field strength is 
\beq
B_i=F_{i4}= {\bf b}_i\cdot {\bf H}.
\eeq
The Pontryagin index  density of this configuration is
\beq 
n_{\rm Pont} = \frac{1}{32 \pi^2 c_2(G)} \tr (F_{\mu\nu}
\tilde{F}_{\mu\nu}) = \frac{{\bf b}_i\cdot {\bf b}_i}{8\pi^2}.
\eeq

Once we choose the symmetric gauge for $B_i$ and the Landau gauge for
$F_{i4}$ so that $\partial_4 A_i=0$, the background field is identical
to those in Eq.~(\ref{back3}) with $A_4=\phi$. Contrast to the
magnetic monopole case, the plane (\ref{plane}), which is now a three
volume, is not special any more as ${\bf h}\cdot \balpha$ can be
changed by the unbroken gauge transformations.  Rather for each ${\bf
b}_i\cdot\balpha$, $B_i$ is the magnetic field through the
two-dimensional plane defined by $x^i{\bf b}_i\cdot\balpha=0$ and
$x^4=0$, and $F_{i4}$ is the magnetic field through the
two-dimensional $x^i\sim {\bf b}_i\cdot\balpha$ and $x^4$ plane.

The linear fluctuation equations for $\delta A_\mu$ will be identical
to Eq.~(\ref{linear3}), once we allow the $x^4$ dependence. The
background gauge condition is then identical to Eq.~(\ref{backg3}).
The fluctuations along the unbroken abelian direction is again
independent of the background and nonrenormalizable. They changes the
parameters ${\bf b}_i $ and ${\bf h}$.

On the other hand, for the fluctuations along a raising operator
$E_\balpha$, we use Eq.(\ref{wand}) to introduce $W_\mu$.  With the
$x^4$ dependence, $e^{ip^4 x^4}$, and a spatial rotation such that
${\bf b}_i\cdot \balpha$ lying along the $z$ direction, the zero mode
solutions are identical to monopole case (\ref{zerom}) once we use
Eq.~(\ref{bandz}) and replace $z_0 $ by $z_0 + p^4/b$. This shows that
the zero mode is not confined on the plane (\ref{plane}).  Rather the
four zero mode solutions are the product of two wave functions of the
lowest Landau level, one on the $x-y$ plane and another on the $z-x^4$
plane. First is written in the symmetric gauge and second is in the
Landau gauge.  As far as these zero modes are concerned, the
translation along the $x$ direction and those along the $y$ direction
do not commute. This makes again the $x-y$ plane a phase space or a
quantum plane with unit quantum cell of area $2\pi/b$. Similarly the
$z-x^4 $ plane becomes a quantum plane of unit quantum cell of area
$2\pi/b$.  Thus on the $R^4$ space the number density of zero modes
for each positive root $\balpha$ is
\beq 
n_\balpha = 4  \left( \frac{{\bf b}_i \cdot \balpha}{2\pi}\right)^2. 
\eeq

The number density of zero modes from all positive roots is the sum
\beq
n_{\rm zero} = \sum_{\balpha>0} n_\balpha
 =  \frac{{\bf b}_i \cdot {\bf b}_i}{2\pi^2} c_2(G),
\eeq
where Eq.~(\ref{identity}) is used.  The number of zero modes per unit
Pontryagin index is then 
\beq
\frac{n_{\rm zero}}{n_{\rm Pont}}= 4\times c_2(G).
\eeq
This number is exactly identical to what is obtained from the index
theorem~\cite{bernard} and also from the consideration of constituent
monopoles of calorons~\cite{klee}.

Instantons appear as the bound states of D0 branes on overlapped D4
branes of the type 2A theory. Our case will be the limit where the
number of D0 branes is infinite. There is a work recently on this
limit~\cite{kraus}. The D-brane picture of our homogeneous solution
for instanton is much less clear than that for monopoles.  While we
assume that homogeneous self-dual configuration can be obtained by
arranging hte infinite number of instantons right, we do not  know how
they are put together for the homogeneous configuration.

In this paper we explored the homogeneous BPS field configuration as
the sheets of infinite number of BPS magnetic monopoles. The
normalizable zero modes is confined at the magnetic monopole sheet,
even though their position on the plane is not localized. Their
position on plane are described by the quantized phase space or
noncommutative two-plane. We showed that the number of normalizable
zero modes per unit magnetic monopole charge is four. For instantons,
the number of normalizable zero modes per unit Pontryagin number is
identical to that of  single instanton without uniform background.
The position space $R^4$ of the normalizable zero modes become
noncommutative four space. We also discussed the D-brane pictures of
these homogeneous configurations.

There are several questions arising naturally.  First question is
whether single magnetic monopole can be separated from the infinite
sheet. Forementioned work on the $SU(2)$ case by C. Lee and Q.H. Park
suggests that it cannot be done without singular behavior in other
side of the monopole sheet. We think this needs a further
clarification.  On the other hand, there exists a remarkable Minkowski
solution~\cite{minkowski} in the instanton case. It describes the
instanton lump in the uniform background.

When ${\bf b}\cdot \balpha =0$ for a root $\balpha$, one can embed the
corresponding $SU(2)$ monopole or instanton solutions.  In the
monopole case with the $SU(N)$ gauge group, two D3 branes are parallel
and any finite number of D string can be inserted. In the instanton
case, we do not know what arrangement of D4 and D0 makes such a
special case. Note that the number of zero modes per unit Pontryagin
index is independent whether ${\bf b}_i\cdot \balpha =0$ or not.

Second, it would be interesting to find out the moduli space dynamics
of zero modes of these monopole and instanton sheet.  The moduli space
would be now infinite dimensional. As the position space of these zero
modes are noncommutative, the moduli space may be noncommutative and
the low energy Lagrangian could be some sort of a field theory of
noncommutative variables.

Third, let us ask how $SU(2)$ monopole or instanton sheets can be
obtained from ADHM and Nahm formalisms~\cite{adhmn}. Since there are
infinite number of monopoles and instantons, the $SU(\infty)$ group
appears naturally.  Ward's work\cite{ward} on Nahm's equation should
be relevant to the magnetic monopole sheet. But we do not know the
corresponding ADHM formalism or the ADHMN construction of solutions.
In addition,  for instanton case, our work seems to be related to the
recent work on the noncommutative geometry~\cite{schwarz}. It remains
to be clarified.

\vspace{4mm}

\centerline{\bf Acknowledgments} 

I thank Bum-Hoon Lee, Choonkyu Lee, Nick Manton, Soo-Jong Rey, and
Erick Weinberg for useful discussions. This work is done partly while
I was visiting the SNU-CTP during the 1998 summer. This work was also
supported by KOSEP 1998 Interdisciplinary Research Program and SRC
program of SNU-CTP, and Ministry of Education BSRI 98-2418.

\end{document}